\documentclass[a4paper]{jpconf}
\usepackage{graphicx}

\usepackage{amssymb}
\usepackage{amsfonts}
\usepackage{amsmath}
\usepackage{amsthm}
\usepackage{epsfig}
\usepackage{color}
\usepackage{bbm}

\begin{document}
\title{Amplitude chimeras and chimera death in dynamical networks}
\author{Anna Zakharova, Marie Kapeller, and Eckehard Sch{\"o}ll$^*$}
\address{Institut f{\"u}r Theoretische Physik, Technische Universit\"at Berlin, Hardenbergstra\ss{}e 36, 10623 Berlin, Germany}
\ead{schoell@physik.tu-berlin.de}
\begin{abstract}
We find chimera states with respect to amplitude dynamics in a network of Stuart-Landau oscillators. These partially coherent and partially incoherent spatio-temporal patterns appear due to the interplay of nonlocal network topology and symmetry-breaking coupling. As the coupling range is increased, the oscillations are quenched, amplitude chimeras disappear and the network enters a symmetry-breaking stationary state. This particular regime is a novel pattern which we call chimera death. It is characterized by the coexistence of spatially coherent and incoherent inhomogeneous steady states and therefore combines the features of chimera state and oscillation death. Additionally, we show two different transition scenarios from amplitude chimera to chimera death. Moreover, for amplitude chimeras we uncover the mechanism of transition towards in-phase synchronized regime and discuss the role of initial conditions. 

\end{abstract}

\section{Introduction}

Chimera states represent an intriguing phenomenon in a dynamical network of coupled elements, which spontaneously separates into two coexisting domains with dramatically different behavior \cite{ZAK14}. They occur, surprisingly, in networks of identical units and symmetric coupling schemes \cite{KUR02a,ABR04}. In spite of the fact that these symmetry-breaking states have been the subject of many investigations, e.g., \cite{SET08,LAI09,MOT10,MAR10,OLM10,BOR10,SHE10,WOL11,OME11,LAI11,OME13,HIZ13}, there still remain many issues to be addressed. A recent review on chimera states \cite{PAN15} discusses a number of significant open questions concerning, for instance, the necessary conditions for their existence or generalizations of chimera states. A generalization to amplitude dynamics and to steady state coherence-incoherence patterns has been provided in \cite{ZAK14}. In particular, in \cite{ZAK14} it has been shown that chimeras of steady states appear due to a special symmetry-breaking 
 coupling which originates 
from oscillation death investigations \cite{BAR85,TSA06,
ULL07,KUZ04}. Oscillation death, i.e., the stabilization of inhomogeneous steady states in coupled oscillator networks, appears when a homogeneous steady state splits into at least two distinct symmetry-breaking branches - upper and lower \cite{ZAK13a}. In comparison with amplitude death, which is another type of oscillation suppression, i.e., the stabilization of an unstable homogeneous steady state, oscillation death is much less investigated \cite{KOS13}.

Initially found for phase oscillators \cite{KUR02a}, chimera states can be treated as long chaotic transients towards the in-phase synchronized regime. In this context the network size plays an essential role. In particular, it has been shown theoretically \cite{WOL11} and confirmed experimentally \cite{ROS14a} that the chimera lifetime grows exponentially with the system size. Therefore, chimera states are stable for infinitely large networks and transient for finite-size ensembles. Recently, in \cite{ASH14} the question of what is the smallest possible number of oscillators allowing for chimera states has been addressed. Another important problem, also considered as a necessary condition for the existence of chimera states, is the choice of initial conditions. Random initial conditions do not always guarantee chimera behavior. This is due to the fact that classical chimera states typically coexist with the completely synchronized regime. The coexistence of two stable solutions is a
  signature of bistability 
which is typically associated with hysteresis. The basin of attraction for chimera states can be relatively small in comparison with that of the synchronized state. For that reason chimera states remained for a long time undetected. Indeed, the first experimental studies \cite{HAG12,TIN12,MAR13,LAR13,SCH14a,WIC13,WIC14} on chimera states were provided only twelve years after their theoretical discovery. Consequently, it is reasonable to use specially prepared initial conditions to ensure chimera patterns. However, it is worth noting that a chimera state is not just a temporary trace of initial conditions which disappears in time, but a persisting pattern with a long lifetime.

In the present work we investigate the connection between two symmetry-breaking effects: chimera states and oscillation death. We show that the interplay of nonlocal coupling topology and symmetry-breaking coupling leads to a number of novel partially coherent inhomogeneous spatial patterns, for instance amplitude chimeras and chimera death. Additionally, we analyze the transition scenarios between these two patterns. Moreover, the transient amplitude chimeras and the transition mechanism from amplitude chimera to a synchronized regime is investigated. Further, the newly found amplitude chimeras are analyzed with respect to initial conditions and transients. In particular, we aim to understand the mechanism of transition from amplitude chimera to in-phase synchronization. Finally, we address the question of how the lifetime of amplitude chimeras depends on initial conditions and what are the optimal initial distributions.

\section{Model}
We analyze the paradigmatic model of Stuart-Landau oscillators \cite{KUR02a,ATA03,FIE09,CHO09,KYR13,SCH13b,POS13a}:
\begin{equation}
  \label{eq:model}
  \dot{z}=f(z) \equiv (\lambda+i\omega - \left|z\right|^2)z,
\end{equation}
where $z = r e^{i \phi}=x+iy \in \mathbbm{C}$, $\lambda, \omega \in \mathbbm{R}$. For $\lambda>0$, the uncoupled system exhibits self-sustained limit cycle oscillations with radius $r_0=\sqrt{\lambda}$ and frequency $\omega$. Therefore, the Stuart-Landau system represents a generic model for nonlinear oscillators close to a Hopf bifurcation. We investigate a ring of $N$ nonlocally 
coupled Stuart-Landau oscillators  \cite{ZAK14}:
\begin{equation}
   \label{eq:network}
   \dot{z}_j = f(z_j) + \frac {\sigma}{2P} \sum_{k=j-P}^{j+P} ( \mathrm{Re} z_k-  \mathrm{Re} z_j),
\end{equation}
where $j=1,2,...,N$.
The coupling parameters, which are identical for all links, are the
coupling strength $\sigma \in \mathbbm{R}$ and the coupling range $P/N$, where $P$ corresponds to the number of nearest neighbors in each direction on a ring. Here we consider coupling only in the real parts, since this breaks the rotational $S^1$ symmetry of the system which is a necessary condition for the existence of nontrivial steady states $z_j \neq 0$ and thus for oscillation death \cite{ZAK13a}. With respect to applications this means that the oscillators are coupled only through a single real variable $x$. 
If compared with slow-fast systems such as, for example, the FitzHugh-Nagumo oscillator, which demonstrates two separated time scales, 
the Stuart-Landau oscillator is a generic model for sinusoidal oscillations which have only one characteristic time scale defined by the frequency $\omega$.

\section{Amplitude chimeras and chimera death states}

\begin{figure}[]
\begin{center}
\includegraphics[width=0.8\textwidth]{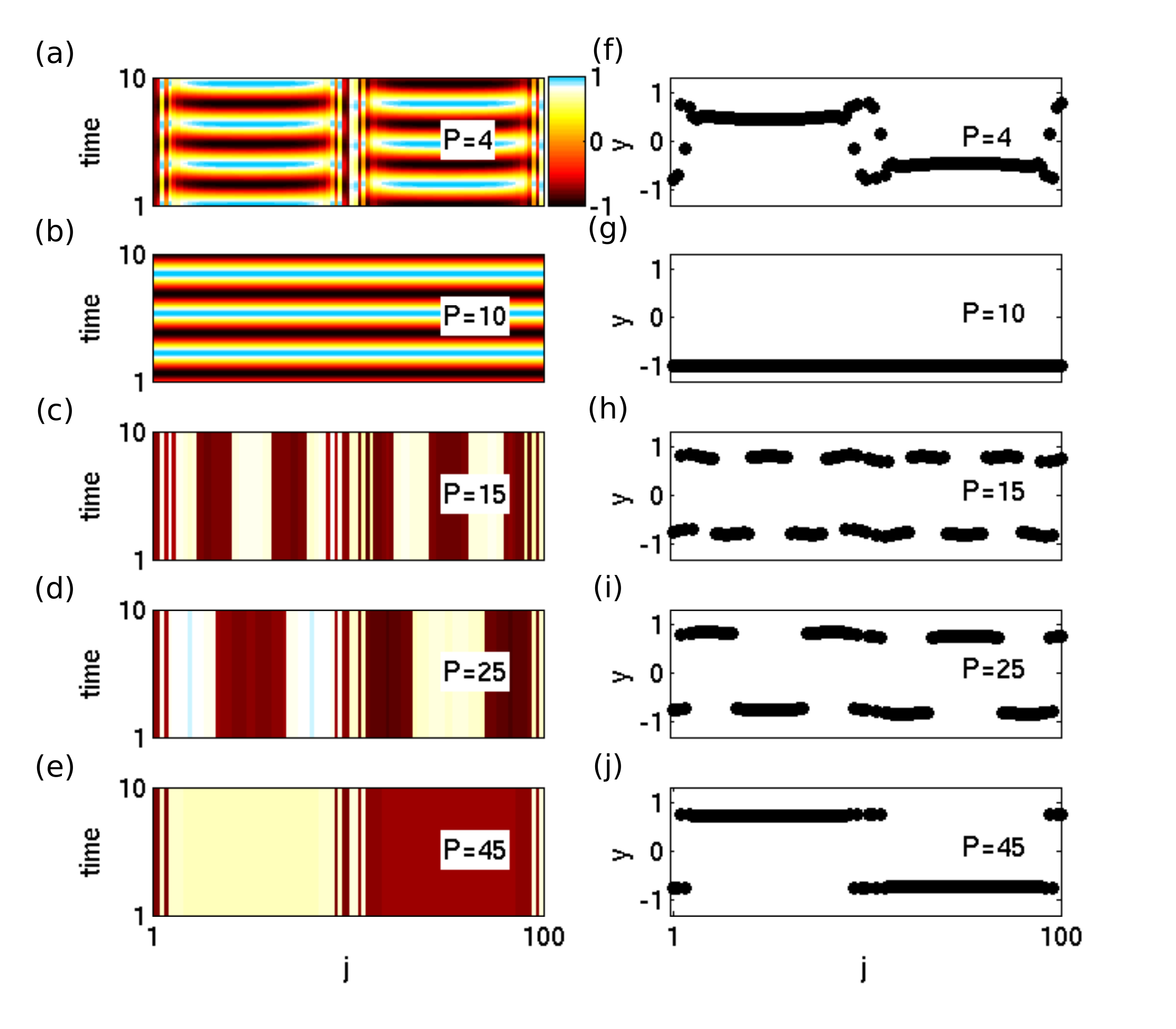}
\end{center}
\caption{Space-time plots (left panel) and snapshots for the variable $y_j(t)$ (right panel) in a network of Stuart-Landau oscillators for coupling strength $\sigma=10$ and varying nearest neighbors number $P$. (a),(f) $P=4$: amplitude chimera; (b),(g) $P=10$: in-phase synchronized oscillations; (c),(h) $P=15$: multi-cluster ($>3$) chimera death; (d),(i) $P=25$: $3$-cluster chimera death; (e),(j) $P=45$: $1$-cluster chimera death. Other parameters: $N=100$, $\lambda=1$, $\omega=2$.}
\label{fig:1}
\end{figure}

\begin{figure}[]
\begin{center}
\includegraphics[width=0.8\textwidth]{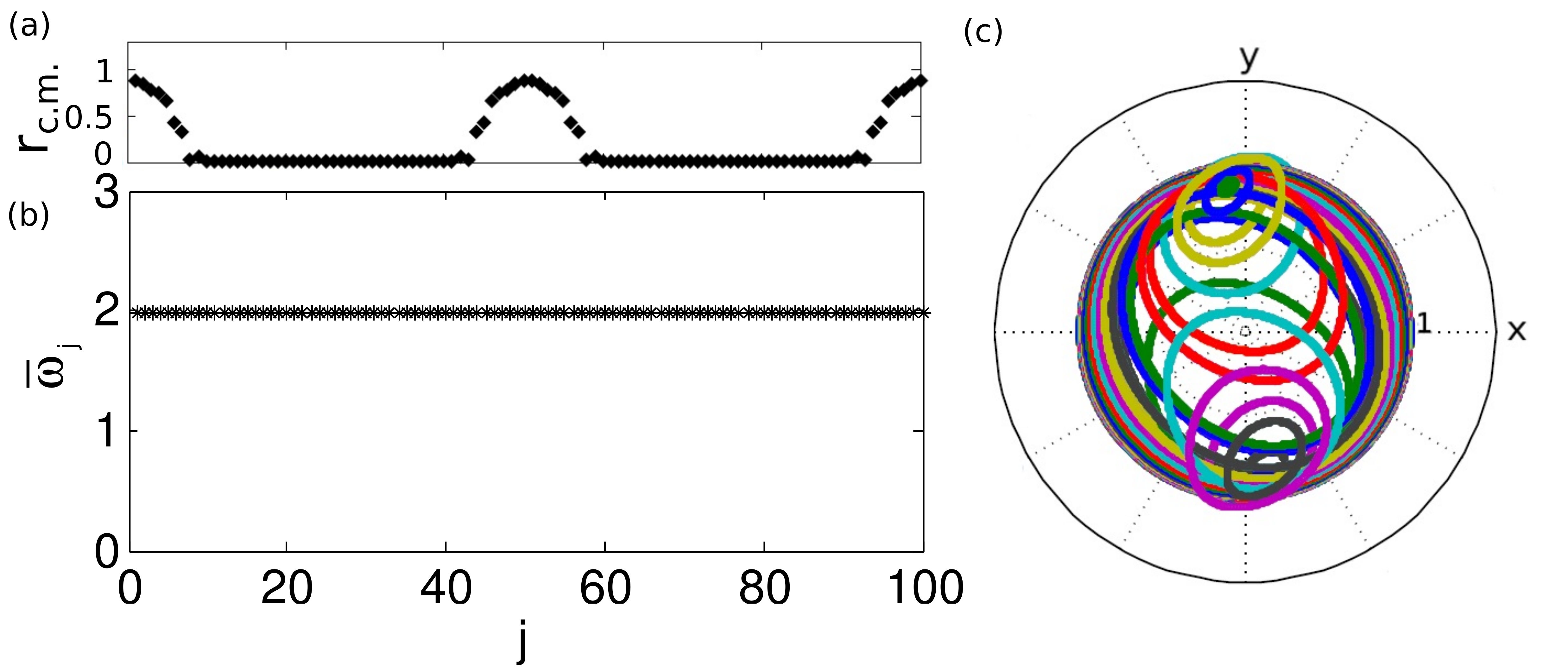} 
\end{center}
\caption{Amplitude chimera: (a) snapshot at $t=1000$ for $r_{c.m.}$ (the distance between the center of mass for every oscillator and the origin); (b) mean phase velocity profile; (e) phase portraits of all oscillators in the complex $z=x+iy$ plane. Parameters: $N=100$, $P=4$, $\sigma=14$, $\lambda=1$, $\omega=2$.}
\label{fig:2_new}
\end{figure}

While tuning the coupling range $P$ for a fixed value of the coupling strength $\sigma$ we uncover a variety of dynamical regimes in equation (2) which are shown as space-time plots, color-coded by the variable $y$, and snapshots in Fig. \ref{fig:1}. In particular, we find chimera behavior with respect to amplitude dynamics, i.e., amplitude chimeras \cite{ZAK14}, where one part of the network is oscillating with spatially coherent amplitude, while the other displays oscillations with spatially incoherent amplitudes (Fig. \ref{fig:1}a,f). It is important to note that within the incoherent domain of the amplitude chimera the center of mass of each oscillator $z_{c.m.}=\int_0^T z_j(t) dt/T$, where $T=2\pi/\omega$ is the oscillation period, is shifted away from the origin, while the elements from the coherent subgroup oscillate around the origin. To illustrate this fact we calculate the distance between the center of mass of each oscillator and the origin $r_{c.m.}$ (Fig. \ref{fig:2_new}
 a). Surprisingly, the 
shape 
of this profile is very similar to that of the mean phase velocity profile for the classical phase chimeras. However, for amplitude chimeras the mean phase velocity profile $\bar{\omega}_j$ is flat (Fig. \ref{fig:2_new}b). This fact discloses the crucial feature of amplitude chimera: averaged phase velocities remain the same for every element of the network, since the phases are correlated even within the incoherent domain. In more detail, 
Fig. \ref{fig:2_new}c shows the phase portraits of all oscillators in the complex $z$ plane clearly demonstrating the limit cycles with different amplitudes and centers of mass. The nodes from the coherent part oscillating with larger amplitudes around the origin perform fast motions in the complex $z=x+iy$ phase plane, while the elements of the incoherent domain with smaller amplitudes are slowed down. Therefore, the angular frequencies of all the nodes on average remain the same. Consequently, we observe pure amplitude chimera - chimera behavior exclusively with respect to amplitude dynamics rather than the phase, in contrast to amplitude-mediated chimeras, for which both phase and amplitude are in a chimera state \cite{SCH14a,SET13}.

\begin{figure}[]
\begin{center}
\includegraphics[width=0.8\textwidth]{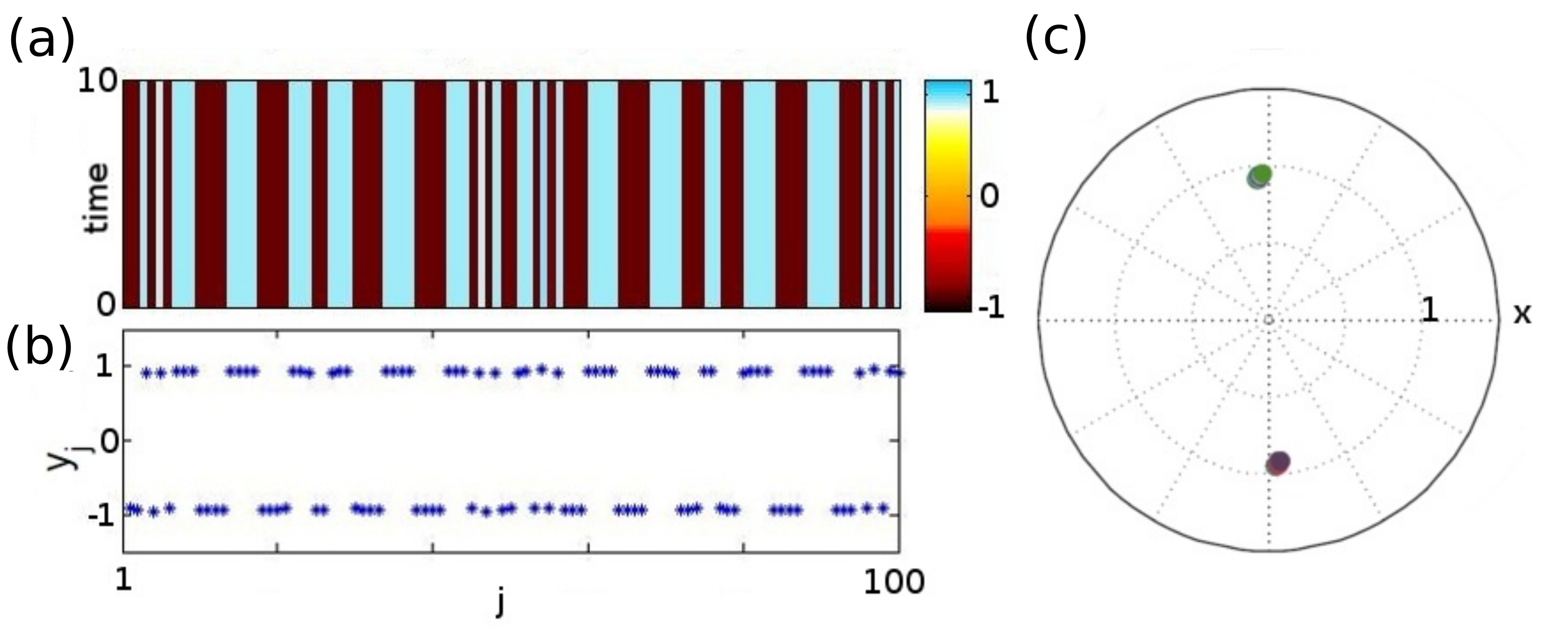}
\end{center}
\caption{Multi-cluster ($>3$) chimera death: (a) space-time plot for the variable $y_j$; (b) snapshot for the variable $y_j$; (c) phase portrait of all oscillators in the complex $z=x+iy$ plane. Parameters: $N=100$, $P=5$, $\sigma=26$, $\lambda=1$, $\omega=2$.}
\label{fig:3_new}
\end{figure}

The increase of the coupling range $P/N$ induces a transition from amplitude chimeras to an in-phase synchronized state (Fig. \ref{fig:1}b,g). By increasing the coupling range even further we detect a novel pattern which provides bridging between chimera states and oscillation death. Therefore, we call it {\it chimera death} \cite{ZAK14} (Fig. \ref{fig:1}c-e,h-j). In this regime the oscillations die out in a peculiar way. The population of identical oscillators breaks up into two domains: (i) spatially coherent oscillation 
death, where the neighboring elements of the network are correlated forming a regular inhomogeneous steady state, and (ii) spatially incoherent oscillation death, where the sequence of populated branches of the inhomogeneous steady state of neighboring nodes is completely random. It is important to note that the term ``coherent/incoherent'' refers to the coherence/incoherence in space, 
i.e., spatial correlation, which should be distinguished from temporal coherence which refers to correlations in time of 
the dynamics. Because of symmetry reasons, for a node on the upper branch $y^{*1} \approx +1$ of the inhomogeneous steady state in the left half of the system, there always exists a mirror state shifted by phase $\pi$ (anti-phase) in the complex plane, i.e., located on the lower branch $y^{*2} \approx -1$, in the right half of the system.  
Interestingly, the increase of the coupling range for fixed coupling strength also induces structural changes of the chimera death pattern: the coherent spatial domain may consist of one or several clusters where the nodes are on the same branch of the inhomogeneous steady state. With increasing coupling range the number of clusters in the coherent spatial domain is decreased, see the scenario in Fig. \ref{fig:1}, from (c,h) (5 clusters) via (d,i) (3 clusters) to (e,j) (1 cluster). The illustration of the chimera death pattern with the maximum number of clusters in the coherent part is provided by Fig. \ref{fig:3_new}a-c. The phase portrait typical for chimera death is shown in Fig. \ref{fig:3_new}c, it consists of two fixed points shifted by a phase $\pi$, corresponding to the two branches $y^{*1}$, $y^{*1}$ of the inhomogeneous steady state. It should be noted that these two fixed points are very close to the centers of mass of those oscillations in Fig. \ref{fig:2_new}c with the smallest amplitude, i.e., 
in the center of the incoherent domain of the amplitude chimera.

\begin{figure}[]
\begin{center}
\includegraphics[width=0.49\textwidth]{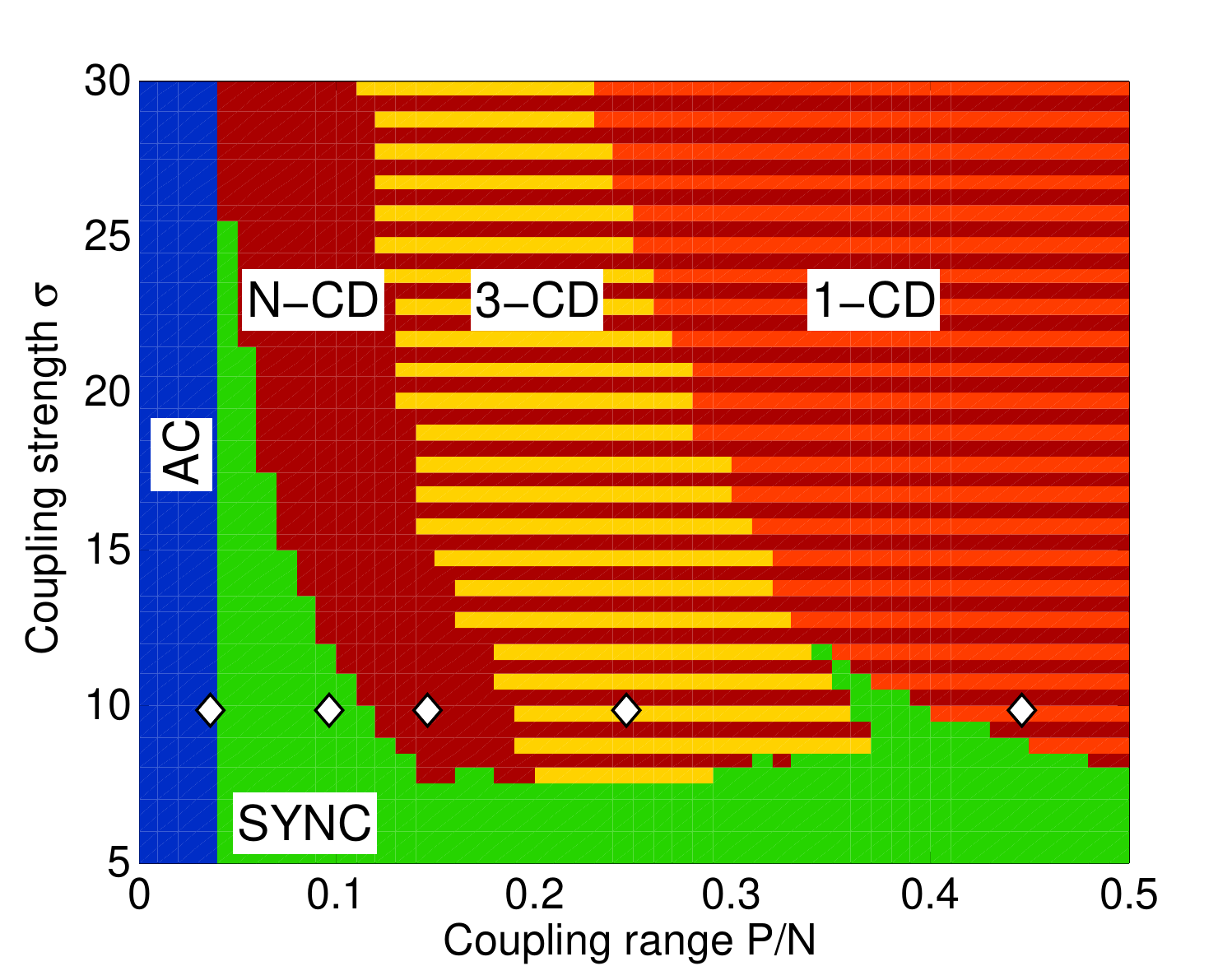}
\end{center}
\caption{Map of dynamical regimes for $N=100$, $\lambda=1$, $\omega=2$ in the plane of coupling range $P/N$ and coupling strength $\sigma$ for specially prepared initial conditions, showing $1$-cluster chimera death ($1$-CD); $3$-cluster chimera death ($3$-CD); multi-cluster ($>3$) chimera death ($N$-CD); amplitude chimera (AC); 
in-phase synchronized oscillations (SYNC). Diamonds mark the parameter values chosen in Fig. \ref{fig:1}. Other parameters: $N=100$, $\lambda=1$, $\omega=2$. The detailed explanation of initial conditions is given in Section 5.}
\label{fig:2}
\end{figure}

To  provide an overall view on the network behavior for the wide range of coupling parameters we plot the map of regimes in the plane of coupling range and coupling strength. The oscillatory behavior of the network is represented by amplitude chimeras (blue region in Fig. \ref{fig:2}), which are observed for small coupling range, and by in-phase synchronized oscillations (light green region in Fig. \ref{fig:2}). The steady state solutions occur for larger values of coupling parameters and are manifested by chimera death (red regions in Fig. \ref{fig:2} with different hatching). The region of chimera death is divided into several regimes depending on the number of coherent clusters. Chimera death may consist of two coherent domains, which are formed by one cluster on the upper and one on the lower branch of the steady state (dark and light red stripes in Fig. \ref{fig:2}). Moreover, the coherent regions may split into three clusters each (dark red and yellow stripes in Fig. \ref{fig:2
 }) or more (dark red in 
Fig. \ref{fig:5}). The typical space-time patterns and snapshots in panels (c,d,e) and (h,i,j) of Fig \ref{fig:1} provide illustrations of the different multi-cluster chimera death states, and the corresponding coupling parameters are indicated in the phase diagram (Fig \ref{fig:2}) by empty diamonds. It is important to note that the chimera death regime is characterized by high multistability. Therefore, chimera death patterns with different number of clusters in the coherent part coexist. The map of regimes strongly depends on initial conditions, in particular, the borders between chimera death patterns with different number of clusters. 
However, the boundary separating the oscillatory regime from the steady state regimes (with multiple cluster numbers) appears to be less sensitive to initial conditions and remains almost the same for different realizations of initial conditions. 
The existence of two distinct transition scenarios from the oscillatory to the steady state regime becomes evident from Fig. \ref{fig:2}.
  For a small value of coupling strength $\sigma=10$ the amplitude chimera gives way to the chimera death state after passing through in-phase synchronized oscillations when the coupling range is increased (diamonds in Fig. \ref{fig:2}). In contrast, for a large value of the coupling strength, for example, $\sigma=26$ a slight increase of the coupling range from $P/N=0.04$ to $P/N=0.05$ destroys amplitude chimeras and directly leads to chimera death.

\section{Transient behavior of amplitude chimeras}

An important feature of classical chimera states is their transient character and strong dependence on initial conditions. Our numerical analysis shows that amplitude chimeras although manifesting a long lifetime (up to $10^4$ dimensionless time units) for the optimal choice of initial conditions, can rapidly transform into in-phase synchronized regime for completely random initial conditions.

\begin{figure}[]
\begin{center}
\includegraphics[width=0.8\textwidth]{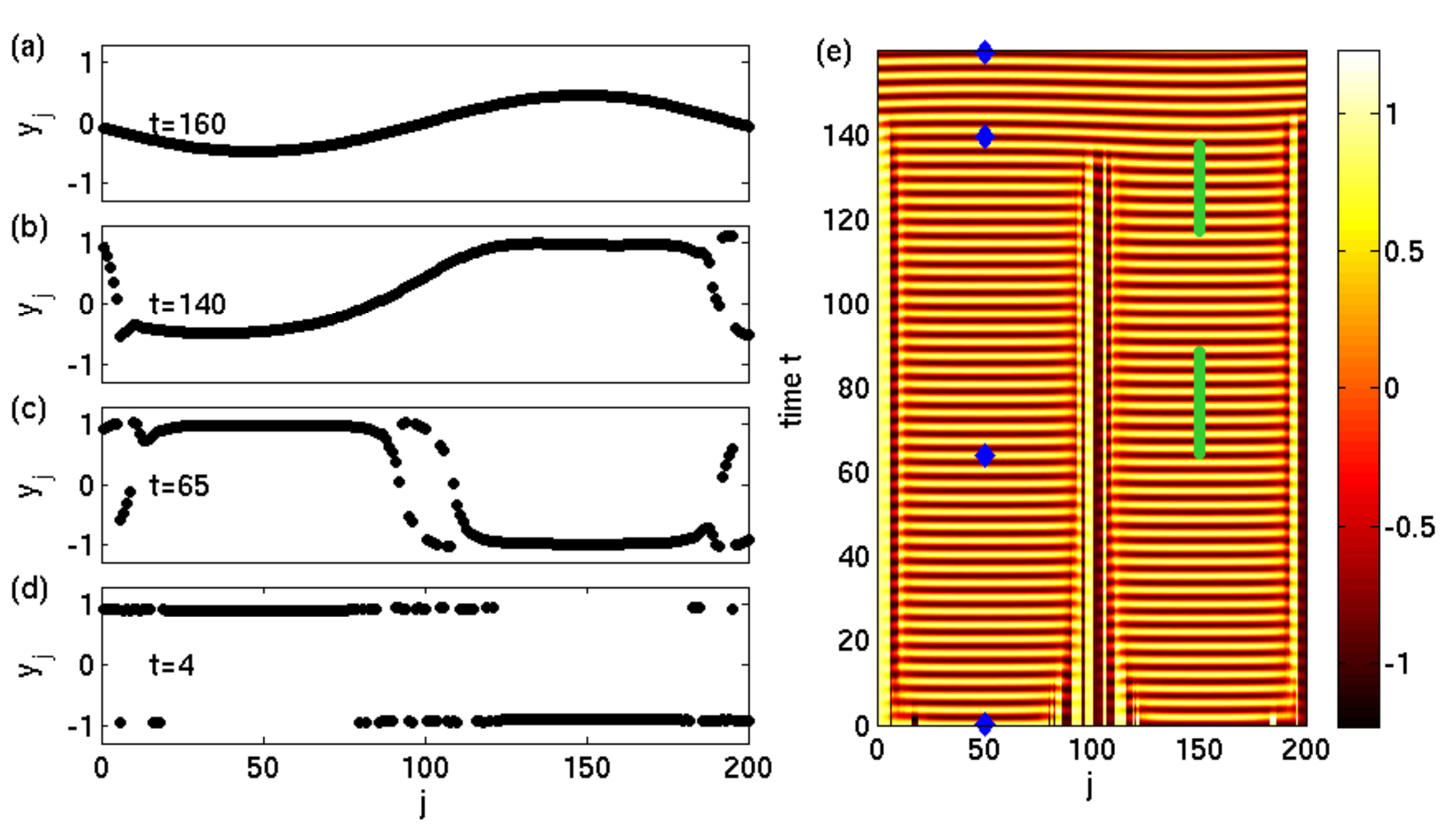}
\end{center}
\caption{Transient amplitude chimera for $N=200$, $\lambda=1$, $\omega=2$, $P=5$ and $\sigma=20$. Snapshots for the variable $y_j$ (left panel): (a) $t=160$: in-phase synchronized oscillations; (b) $t=140$: deformed amplitude chimera at the transition point; (c) $t=65$: amplitude chimera; (d) $t=4$: trace of initial condition. The right panel shows the space-time plot for the variable $y_j(t)$. Blue diamonds mark the time values chosen for the snapshots in the left panel. The green lines mark the time ranges chosen for the phase portraits in Fig. \ref{fig:4}.}
\label{fig:3}
\end{figure}

To gain more insight into the transient behavior of amplitude chimeras we trace the system's evolution in time having fixed all other parameters. Figure \ref{fig:3}a-d shows corresponding snapshots for different time values. As initial condition here we take chimera death ($t=0$). The pattern at an early stage ($t=4$) shown in Fig. \ref{fig:3}d is reminiscent of the initial conditions, while the regime of amplitude chimeras (Fig. \ref{fig:3}c) is reached later ($t=65$). However, after approximately $125$ time steps it starts deforming (Fig. \ref{fig:3}b) until it turns into an in-phase synchronized pattern (Fig. \ref{fig:3}a). This is also visible in the
space-time plot (Fig. \ref{fig:3}e).

\begin{figure}[]
\begin{center}
\includegraphics[width=0.8\textwidth]{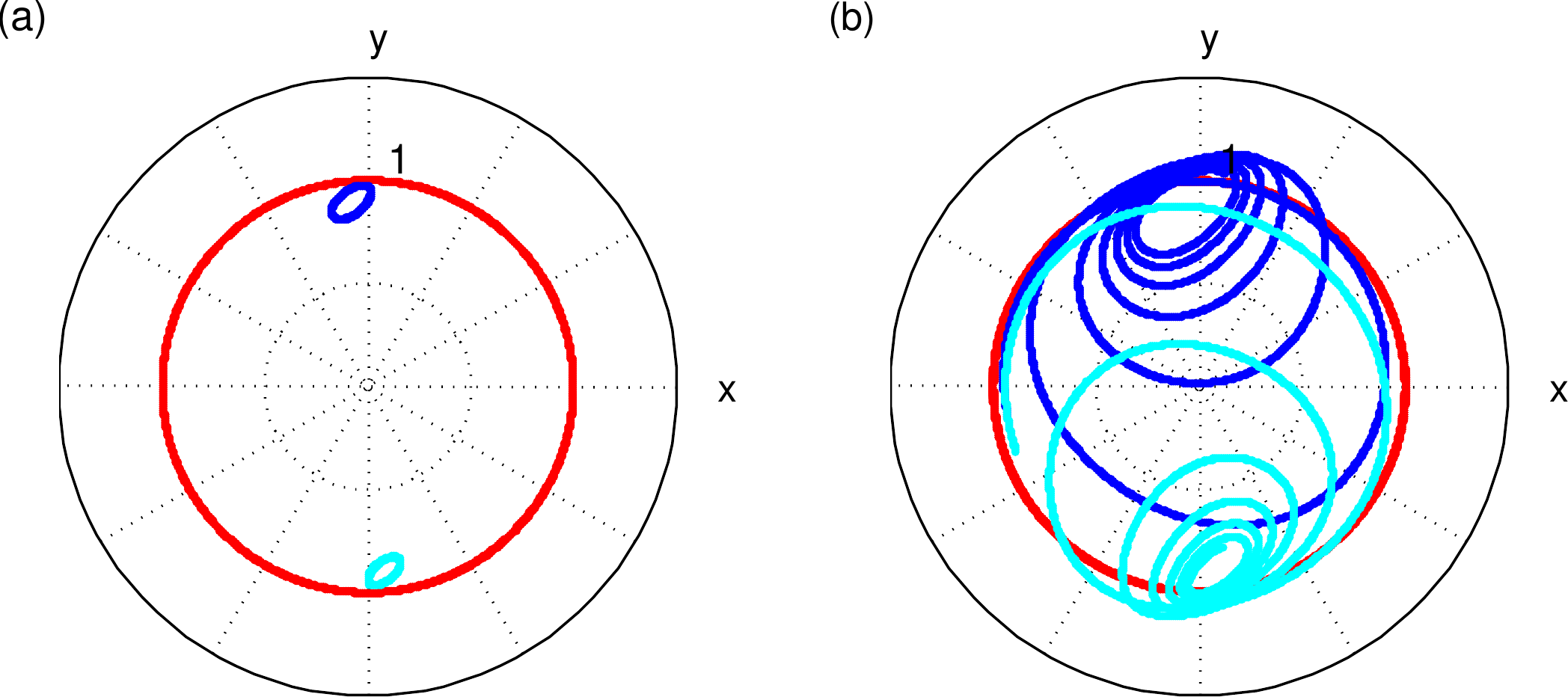}
\end{center}
\caption{Phase portraits of three selected nodes of the network in the complex $z=x+iy$ plane for (a) amplitude chimera state and (b) during the transition to in-phase synchronization. Red color marks the trajectory of node $j=50$ (from the coherent part). Dark and light blue colors show trajectories of nodes from the incoherent domain $j=99$ and $j=102$ correspondingly. Time ranges as marked in green in Fig. \ref{fig:3}. Parameters: $N=200$, $\lambda=1$, $\omega=2$, $P=5$, $\sigma=20$.}
\label{fig:4}
\end{figure}

To disclose the transition mechanism from amplitude chimera to the in-phase synchronized regime we analyze phase portraits in the amplitude chimera state (Fig. \ref{fig:4}a) and during the transition (Fig. \ref{fig:4}b) for three selected nodes of the network: one from the coherent part ($j=50$) and the other two from the incoherent domain ($j=99,102$). The time ranges covered in Fig. \ref{fig:4}a and b, respectively, are indicated by two green vertical lines in the space-time plot of Fig. \ref{fig:3}e. In the regime of amplitude chimera the node from the coherent part ($j=50$) oscillates with large amplitude around the origin, while the two nodes in the incoherent domain ($j=99,102$) oscillate with much smaller amplitudes around their own centers of mass shifted from the origin in two opposite directions (Fig. \ref{fig:4}a). During the transition the amplitudes of the incoherent oscillations grow till they reach the value of the coherent oscillations. At the same time the centers of
  mass for nodes from 
the incoherent domain move towards the origin (Fig. \ref{fig:4}b). Therefore, during the transition the phase trajectories for the nodes $j=99,102$ have spiral shapes (Fig. \ref{fig:4}b) .

An important observation is that the transition mechanism described above depends on the initial conditions. For example, asymmetric initial conditions may lead to another scenario which involves an asymmetric state. In this case nodes from the incoherent part join one by one their neighboring synchronized state. 

\section{The role of initial conditions}
\begin{figure}[]
\begin{center}
\includegraphics[width=0.8\textwidth]{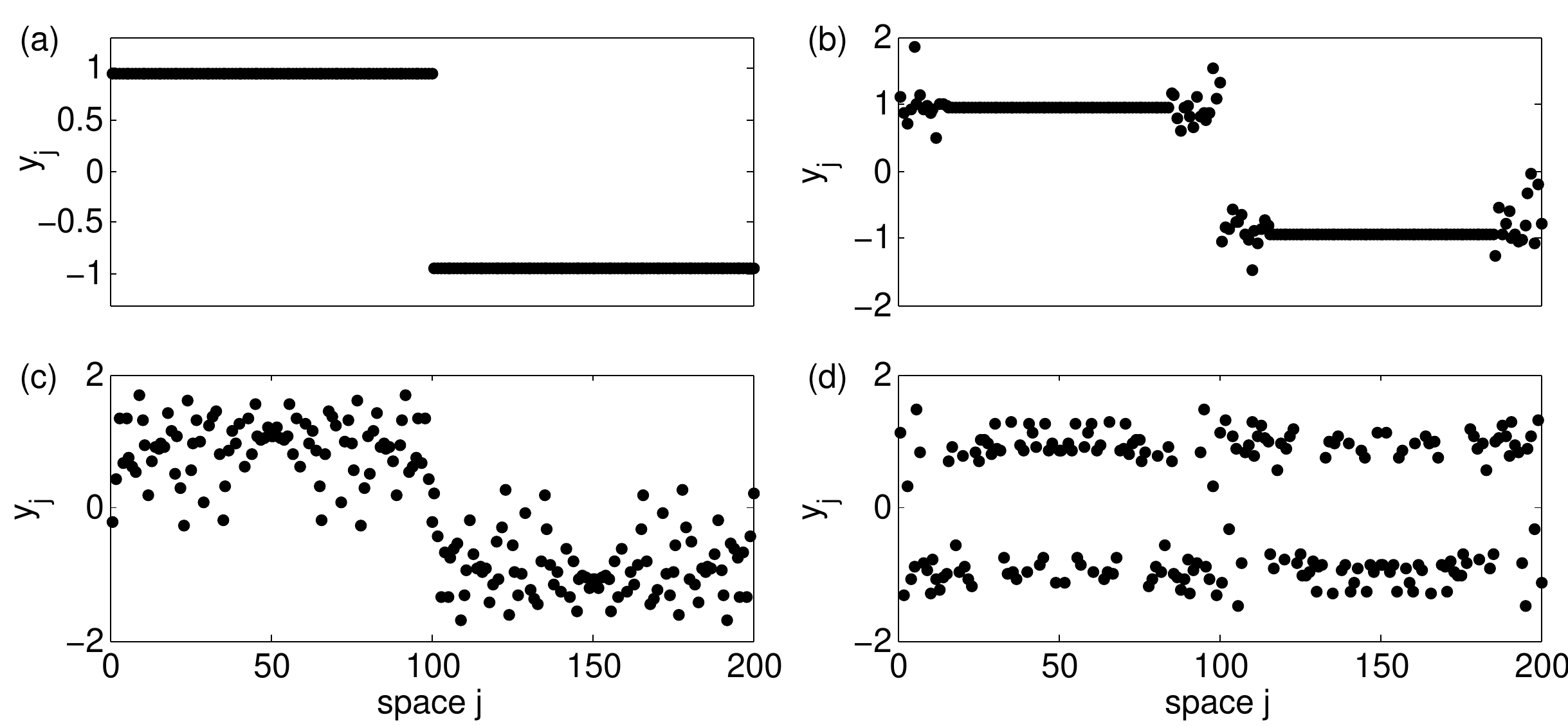}
\end{center}
\caption{Initial conditions for amplitude chimeras used for $N=200$, $\lambda=1$, $\omega=2$, $P=4$ and $\sigma=20$: (a) coherent distribution: one half of the nodes on the upper and one half on the lower branch of the inhomogeneous steady state (amplitude chimera lifetime $t=130$); (b) combination of coherent and incoherent domains with no symmetries in the randomly defined incoherent part (amplitude chimera lifetime $t=25$); (c),(d) random initial conditions with symmetries (optimal amplitude chimera lifetime $t>3000$).}
\label{fig:5}
\end{figure}

The first initial condition we choose originates from the oscillation death problem and is typically used for the inhomogeneous steady state \cite{ZAK13a}. Specifically, the system is divided into two equal domains: one is located on the upper branch and the other is on the lower branch. In our particular example of $N=200$ nodes this implies that one half of the network ($j=1,...,100$) is on the upper branch $y^{*1} \approx +1$ while the other half is on the lower branch $y^{*2} \approx -1$ (Fig. \ref{fig:5}a). In this case the lifetime of amplitude chimeras is $t<200$ and decreases significantly when some small random shift from the branches is applied for every node.

Another type of initial condition we have tested is the modification of the previously discussed set by adding two incoherent regions, where the nodes are distributed randomly (Fig. \ref{fig:5}b). This leads to dramatically short lifetime ($t<30$) of the amplitude chimera, which rapidly transforms into an in-phase synchronized state or a traveling wave. 

Next we allow randomness in initial conditions for all nodes of the network, however, observing certain symmetries. In particular, in Fig. \ref{fig:5}c the positions for $1/4$ of the nodes $j=1,...,50$ are chosen randomly around the upper branch, then these values are mirrored to the nodes $j=51,...,100$ by setting $z_j=z_{1-j+N/2}$.  Finally, the positions of the other half of the nodes $j=101,...,200$ are obtained by applying a phase shift of $\pi$ which we call ``anti-phase partner'' condition: $z_j=-z_{j+N/2}$ with $j$ $mod$ $N$. The initial condition scheme shown in Fig. \ref{fig:5}d is constructed in a similar way as in Fig. \ref{fig:5}c. The only difference is that the values for the first $j=1,...,50$ nodes are chosen randomly around both branches of the inhomogeneous steady state (and not only around the upper branch like in Fig. \ref{fig:5}c). This amendment does not have any impact on the lifetime of amplitude chimeras, which is on the other hand strongly affected by the p
 resence of symmetries in 
the initial conditions. Our simulations suggest that amplitude chimeras tend to have very short lifetime if in the incoherent domain the initial positions of nodes near the upper branch are uncorrelated with those close to the lower branch. An increase of the amplitude chimera lifetime can be achieved by introducing symmetries in the initial distribution (``anti-phase partner'' condition). Therefore, optimal results with $t>3000$ are obtained for random initial conditions with symmetries (Fig. \ref{fig:5}c,d). In conclusion, our numerical analysis shows that for random distributions without symmetries amplitude chimeras appear to be short transients towards in-phase synchronized regime, while symmetric conditions significantly increase their lifetime up to $t>10^4$. Moreover, random initial distributions may also lead to asymmetric spatio-temporal patterns.

\begin{figure}[]
\begin{center}
\includegraphics[width=0.49\textwidth]{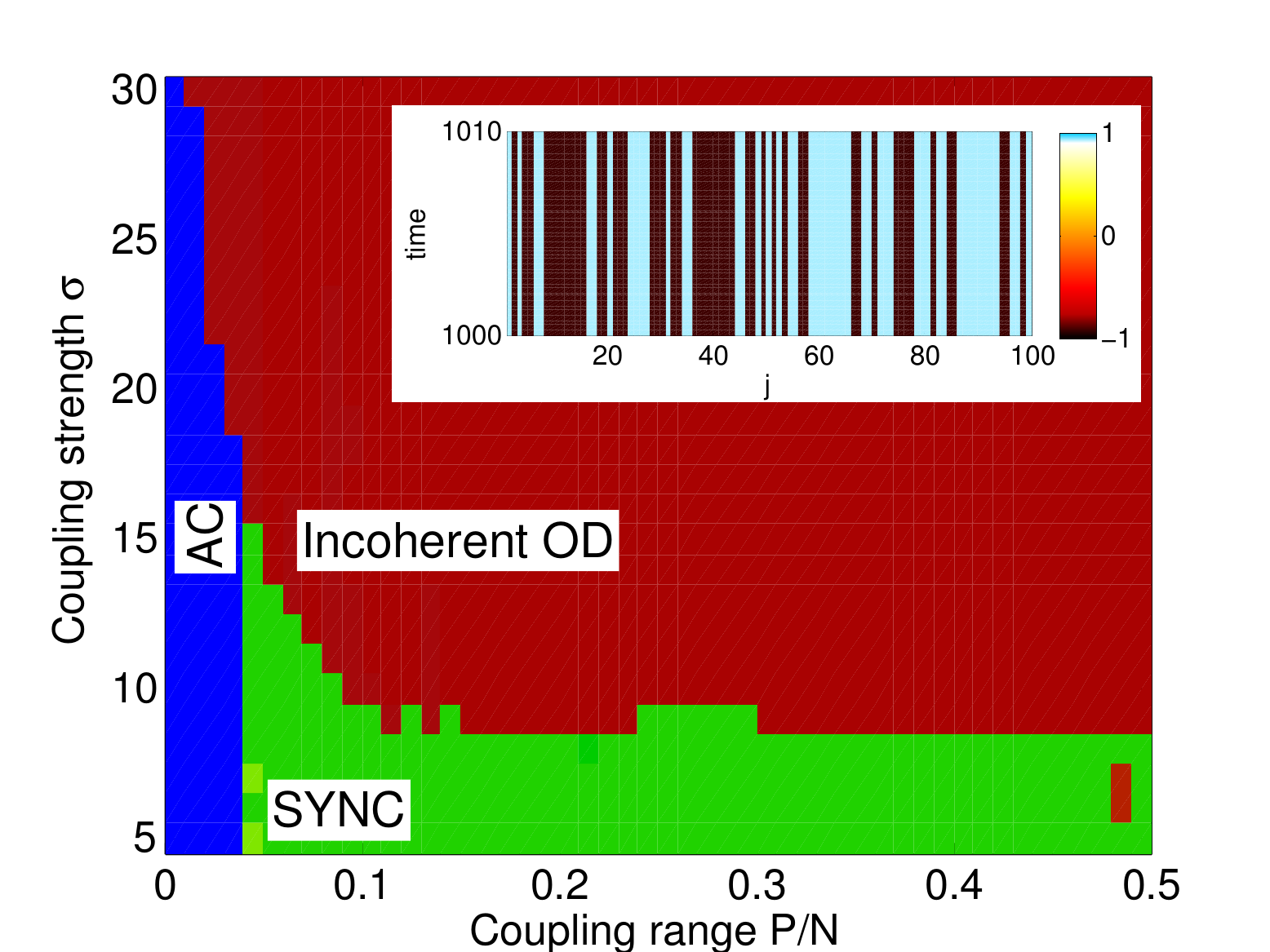}
\end{center}
\caption{Map of dynamical regimes for $N=100$, $\lambda=1$, $\omega=2$ in the plane of coupling range $P/N$ and coupling strength $\sigma$ for random initial conditions as in Fig. \ref{fig:5}(d): amplitude chimera (AC); in-phase synchronized oscillations (SYNC); incoherent oscillation death (Incoherent OD). The inset shows a space-time plot for the variable $y_j(t)$ for the coupling strength $\sigma=30$ and coupling range $P/N=0.5$.}
\label{fig:6}
\end{figure}

To provide a comprehensive view on the impact of initial conditions we compare the map of regimes calculated for a specially prepared initial set (Fig. \ref{fig:2}) and for random initial conditions (Fig. \ref{fig:6}). For the diagram shown in Fig. \ref{fig:2} 
we use as initial condition an amplitude chimera profile at a fixed time (similar to Fig. \ref{fig:3}c), which is obtained in the following way: For a fixed set of parameters in the amplitude chimera regime, the system is divided into two equal domains; half of the nodes are located on the upper branch and the other half is on the lower branch of the inhomogeneous steady state (similar to Fig. \ref{fig:5}a); this initial condition then evolves into an amplitude chimera state which is used as initial condition for all other parameter values $(\sigma, P/N)$. 
The random initial distribution (similar to Fig. \ref{fig:5}d) leads to a significantly different phase diagram (Fig. \ref{fig:6}) in comparison with the specially prepared initial conditions (Fig. \ref{fig:2}). The domain of the amplitude chimera is decreased for strong coupling if compared with Fig. \ref{fig:2}. Moreover, the regime of chimera death obtained from carefully chosen initial conditions (Fig. \ref{fig:2}) is replaced by the regime of incoherent oscillation death (inset in Fig. \ref{fig:6}) which results from random initial conditions. It should be noted that the map of regimes shown in Fig. \ref{fig:6} is obtained for one single realization of random initial distribution and no averaging of initial conditions has been provided. Therefore, it is not statistically representative that chimera death is not observed in Fig. \ref{fig:6}. This is merely due to the fact that the probability to achieve chimera patterns starting from one random realization of the initial conditio
 n is very low. However, 
the occurrence of chimeras for random initial conditions, in general, is not excluded. 
\section{Conclusions}
In conclusion, we have provided a connection between two symmetry-breaking effects - chimera states and oscillation death. In particular, we uncover new spatio-temporal patterns, i.e., amplitude chimeras and chimera death. Amplitude chimeras represent a generalization of chimera behavior to amplitude dynamics, and chimera death generalizes chimeras to steady states. It is shown that different transition scenarios from amplitude chimeras to chimera death are possible. 
Chimera death patterns with different numbers of clusters in the coherent part exhibit a high degree of multistability, which is related to a variety of hysteretic scenarios. Moreover, the transient dynamics of amplitude chimera is investigated disclosing the transition mechanism from amplitude chimera to the in-phase synchronized regime. Additionally, we find optimal initial conditions for amplitude chimeras showing that random distributions with particular symmetries essentially enlarge their lifetime.  

\section*{Acknowledgments}
This work was supported by DFG in the framework of SFB 910.


\clearpage

\section*{References}   
\begin{thebibliography}{10}
\expandafter\ifx\csname url\endcsname\relax
  \def\url#1{{\tt #1}}\fi
\expandafter\ifx\csname urlprefix\endcsname\relax\def\urlprefix{URL }\fi

\bibitem{ZAK14}
A.~Zakharova, M.~Kapeller, and E.~Sch{\"o}ll: {\em Chimera death: Symmetry
  breaking in dynamical networks\/}, Phys.~Rev.~Lett. {\bf 112}, 154101 (2014).

\bibitem{KUR02a}
Y.~Kuramoto and D.~Battogtokh: {\em {Coexistence of Coherence and Incoherence
  in Nonlocally Coupled Phase Oscillators.}\/}, Nonlin. Phen. in Complex Sys.
  {\bf 5}, 380 (2002).

\bibitem{ABR04}
D.~M. Abrams and S.~H. Strogatz: {\em Chimera states for coupled
  oscillators\/}, Phys.~Rev.~Lett. {\bf 93}, 174102 (2004).

\bibitem{SET08}
G.~C. Sethia, A.~Sen, and F.~M. Atay: {\em Clustered chimera states in
  delay-coupled oscillator systems\/}, Phys.~Rev.~Lett. {\bf 100}, 144102
  (2008).

\bibitem{LAI09}
C.~R. Laing: {\em The dynamics of chimera states in heterogeneous {K}uramoto
  networks\/}, Physica D {\bf 238}, 1569 (2009).

\bibitem{MOT10}
A.~E. Motter: {\em Nonlinear dynamics: Spontaneous synchrony breaking\/},
  Nature Physics {\bf 6}, 164 (2010).

\bibitem{MAR10}
E.~A. Martens, C.~R. Laing, and S.~H. Strogatz: {\em Solvable model of spiral
  wave chimeras\/}, Phys. Rev. Lett. {\bf 104}, 044101 (2010).

\bibitem{OLM10}
S.~Olmi, A.~Politi, and A.~Torcini: {\em Collective chaos in pulse-coupled
  neural networks\/}, Europhys. Lett. {\bf 92}, 60007 (2010).

\bibitem{BOR10}
G.~Bordyugov, A.~Pikovsky, and M.~G. Rosenblum: {\em Self-emerging and
  turbulent chimeras in oscillator chains\/}, Phys. Rev. E {\bf 82}, 035205
  (2010).

\bibitem{SHE10}
J.~H. Sheeba, V.~K. Chandrasekar, and M.~Lakshmanan: {\em Chimera and globally
  clustered chimera: Impact of time delay\/}, Phys. Rev. E {\bf 81}, 046203
  (2010).

\bibitem{WOL11}
M.~Wolfrum and O.~E. Omel'chenko: {\em Chimera states are chaotic
  transients\/}, Phys. Rev. E {\bf 84}, 015201 (2011).

\bibitem{OME11}
I.~Omelchenko, Y.~L. Maistrenko, P.~H{\"o}vel, and E.~Sch{\"o}ll: {\em Loss of
  coherence in dynamical networks: spatial chaos and chimera states\/}, Phys.
  Rev. Lett. {\bf 106}, 234102 (2011).

\bibitem{LAI11}
C.~R. Laing: {\em {Fronts and bumps in spatially extended Kuramoto
  networks}\/}, Physica D {\bf 240}, 1960 (2011).

\bibitem{OME13}
I.~Omelchenko, O.~E. Omel'chenko, P.~H{\"o}vel, and E.~Sch{\"o}ll: {\em When
  nonlocal coupling between oscillators becomes stronger: patched synchrony or
  multichimera states\/}, Phys. Rev. Lett. {\bf 110}, 224101 (2013).

\bibitem{HIZ13}
J.~Hizanidis, V.~Kanas, A.~Bezerianos, and T.~Bountis: {\em Chimera states in
  networks of nonlocally coupled hindmarsh-rose neuron models\/}, Int. J.
  Bifurcation Chaos {\bf 24}, 1450030 (2014).

\bibitem{PAN15}
M.~J. Panaggio and D.~M. Abrams: {\em Chimera states: Coexistence of coherence
  and incoherence in networks of coupled oscillators\/}, Nonlinearity {\bf 28},
  R67 (2015).

\bibitem{BAR85}
K.~Bar-Eli: {\em On the stability of coupled chemical oscillators\/}, Physica~D
  {\bf 14}, 242 (1985).

\bibitem{TSA06}
K.~Tsaneva-Atanasova, C.~L. Zimliki, R.~Bertram, and A.~Sherman: {\em Diffusion
  of calcium and metabolites in pancreatic islets: Killing oscillations with a
  pitchfork\/}, Biophys. J. {\bf 90}, 3434 (2006).

\bibitem{ULL07}
E.~Ullner, A.~Zaikin, E.~Volkov, and J.~Garc{\'i}a-Ojalvo: {\em Multistability
  and clustering in a population of synthetic genetic oscillators via
  phase-repulsive cell-to-cell communication\/}, Phys. Rev. Lett. {\bf 99},
  148103 (2007).

\bibitem{KUZ04}
A.~Kuznetsov, M.~Kaern, and N.~Kopell: {\em Synchrony in a population of
  hysteresis-based genetic oscillators\/}, SIAM J. Appl. Math. {\bf 65}, 392
  (2004).

\bibitem{ZAK13a}
A.~Zakharova, I.~Schneider, Y.~N. Kyrychko, K.~B. Blyuss, A.~Koseska,
  B.~Fiedler, and E.~Sch{\"o}ll: {\em Time delay control of symmetry-breaking
  primary and secondary oscillation death\/}, Europhys. Lett. {\bf 104}, 50004
  (2013).

\bibitem{KOS13}
A.~Koseska, E.~Volkov, and J.~Kurths: {\em Oscillation quenching mechanisms:
  Amplitude vs. oscillation death\/}, Phys. Rep. {\bf 531}, 173 (2013).

\bibitem{ROS14a}
D.~P. Rosin, D.~Rontani, N.~D. Haynes, E.~Sch{\"o}ll, and D.~J. Gauthier: {\em
  Transient scaling and resurgence of chimera states in coupled {B}oolean phase
  oscillators\/}, Phys.~Rev.~ E {\bf 90}, 030902(R) (2014).

\bibitem{ASH14}
P.~Ashwin and O.~Burylko: {\em Chimera states in minimal networks of coupled
  phase oscillators\/}, arXiv {\bf 1407.8070} (2014),
  http://arxiv.org/abs/1407.8070.

\bibitem{HAG12}
A.~M. Hagerstrom, T.~E. Murphy, R.~Roy, P.~H{\"o}vel, I.~Omelchenko, and
  E.~Sch{\"o}ll: {\em Experimental observation of chimeras in coupled-map
  lattices\/}, Nature Physics {\bf 8}, 658 (2012).

\bibitem{TIN12}
M.~R. Tinsley, S.~Nkomo, and K.~Showalter: {\em Chimera and phase cluster
  states in populations of coupled chemical oscillators\/}, Nature Physics {\bf
  8}, 662 (2012).

\bibitem{MAR13}
E.~A. Martens, S.~Thutupalli, A.~Fourri{\`e}re, and O.~Hallatschek: {\em
  Chimera states in mechanical oscillator networks\/}, Proc. Nat. Acad.
  Sciences {\bf 110}, 10563 (2013).

\bibitem{LAR13}
L.~Larger, B.~Penkovsky, and Y.~L. Maistrenko: {\em Virtual chimera states for
  delayed-feedback systems\/}, Phys. Rev. Lett. {\bf 111}, 054103 (2013).

\bibitem{SCH14a}
L.~Schmidt, K.~Sch{\"o}nleber, K.~Krischer, and V.~Garcia-Morales: {\em
  Coexistence of synchrony and incoherence in oscillatory media under nonlinear
  global coupling\/}, Chaos {\bf 24}, 013102 (2014).

\bibitem{WIC13}
M.~Wickramasinghe and I.~Z. Kiss: {\em Spatially organized dynamical states in
  chemical oscillator networks: Synchronization, dynamical differentiation, and
  chimera patterns\/}, PLoS ONE {\bf 8}, e80586 (2013).

\bibitem{WIC14}
M.~Wickramasinghe and I.~Z. Kiss: {\em Spatially organized partial
  synchronization through the chimera mechanism in a network of electrochemical
  reactions\/}, Phys. Chem. Chem. Phys.  (2014).

\bibitem{ATA03}
F.~M. Atay: {\em Distributed delays facilitate amplitude death of coupled
  oscillators\/}, Phys.~Rev.~Lett. {\bf 91}, 094101 (2003).

\bibitem{FIE09}
B.~Fiedler, V.~Flunkert, P.~H{\"o}vel, and E.~Sch{\"o}ll: {\em Delay
  stabilization of periodic orbits in coupled oscillator systems\/}, Phil.
  Trans.~R. Soc.~A {\bf 368}, 319 (2010).

\bibitem{CHO09}
C.~U. Choe, T.~Dahms, P.~H{\"o}vel, and E.~Sch{\"o}ll: {\em Controlling
  synchrony by delay coupling in networks: from in-phase to splay and cluster
  states\/}, Phys. Rev.~E {\bf 81}, 025205(R) (2010).

\bibitem{KYR13}
Y.~N. Kyrychko, K.~B. Blyuss, and E.~Sch{\"o}ll: {\em Amplitude and phase
  dynamics in oscillators with distributed-delay coupling\/}, Phil. Trans.~R.
  Soc.~A {\bf 371}, 20120466 (2013).

\bibitem{SCH13b}
I.~Schneider: {\em Delayed feedback control of three diffusively coupled
  {S}tuart-{L}andau oscillators: a case study in equivariant {H}opf
  bifurcation\/}, Phil. Trans.~R. Soc.~A {\bf 371}, 20120472 (2013).

\bibitem{POS13a}
C.~M. Postlethwaite, G.~Brown, and M.~Silber: {\em Feedback control of unstable
  periodic orbits in equivariant hopf bifurcation problems\/}, Phil. Trans.~R.
  Soc.~A {\bf 371}, 20120467 (2013).

\bibitem{SET13}
G.~C. Sethia, A.~Sen, and G.~L. Johnston: {\em Amplitude-mediated chimera
  states\/}, Phys. Rev. E {\bf 88}, 042917 (2013).

\end{thebibliography}
\end{document}